# Reduced Glass Transition Temperature with Preserved Super-Arrhenius Dynamics in Nanoconfined Glass-Forming Nematic

Aleksandra Deptuch[1,*], Eugeniusz Rajewski[2], Ewa Juszyńska-Gałązka[1,3]

[1] The Henryk Niewodniczański Institute of Nuclear Physics Polish Academy of Sciences, Radzikowskiego 152, PL-31342 Kraków, Poland

[2] Faculty of Physics and Applied Computer Science, AGH University of Science and Technology, Reymonta 19, PL-30059 Kraków, Poland

[3] Research Center for Thermal and Entropic Science, Graduate School of Science, Osaka University, 560-0043 Osaka, Japan

*corresponding author, aleksandra.deptuch@ifj.edu.pl

## Abstract

The molecular dynamics and glass-forming behavior of 4-cyano-3-fluorophenyl 4-butylbenzoate (4CFPB) liquid crystal was investigated by broadband dielectric spectroscopy in anodic aluminum oxide cylindrical nanopores of 35 and 73 nm diameter. The primary α-relaxation was analyzed using the Vogel-Fulcher-Tammann (VFT) and Mauro-Yue-Ellison-Gupta-Allan (MYEGA) models. Confinement reduces the glass transition temperature from 217 K in bulk to 208-210 K, while preserving the super-Arrhenius character of the α-relaxation and a fragility index of approximately 60. Unlike in previously studied 6 and 8 nm pores, the secondary β-relaxation remains detectable, with possible intramolecular contributions examined by density functional theory calculations. Next to the pore-size-dependent changes in molecular dynamics and glass transition temperature, cold crystallization is not observed. These findings highlight the distinct effects of geometrical confinement on molecular relaxation, glass formation, and crystallization kinetics.

## 1. Introduction

Geometrical confinement provides an effective approach to modifying the phase behavior and molecular dynamics of liquid crystals, serving as an alternative or additional factor to changing their chemical structure [1-3] and the formulation of mixtures [4-6]. The physical properties can be controlled by incorporating nanoparticles [7,8] or nanotubes [9,10], as well as by embedding liquid crystal compounds in fibers [11] or nanoporous matrices [12-14]. In nanoporous systems, the restricted spatial dimensions and interactions between molecules and pore walls may modify molecular mobility, orientational ordering, phase diagrams, and crystallization kinetics [15,16]. These effects are particularly relevant for glass-forming liquid crystals, where molecular relaxation and structural ordering play important roles in the formation of partially ordered glasses [17,18].

Broadband dielectric spectroscopy (BDS) provides insight into molecular dynamics over a wide range of relaxation times and temperatures [19]. The complex dielectric permittivity reflects the response of the material to an external alternating electric field, allowing molecular relaxation processes and other dielectric contributions to be distinguished. The dielectric spectra can be analyzed using the Cole-Cole model (Equation (1)) [20]:

$$\varepsilon^*(f) = \varepsilon'(f) - i\varepsilon''(f) = \varepsilon_\infty + \sum_j \frac{\Delta\varepsilon_j}{1+(2\pi i f \tau_j)^{1-a_j}} - \frac{iS_1}{(2\pi f)^{n_1}} + \frac{S_2}{(2\pi f)^{n_2}}, \tag{1}$$

where: $\varepsilon_\infty$ is $\varepsilon'$ at frequency $f \to \infty$, $\Delta\varepsilon_j$, $\tau_j$, $a_j$ are dielectric strength, relaxation time, and shape parameter that characterize each relaxation process, while the remaining terms account for ionic conductivity and contributions from relaxation processes located outside the experimental frequency window.

The glass transition temperature $T_g$ is usually defined as the temperature where the structural α-relaxation time $\tau_\alpha$ is equal to 100 s [21]. The temperature dependence of $\tau_\alpha$ can be described using the Vogel-Fulcher-Tammann (VFT, Equation (2)) [21] or Mauro-Yue-Ellison-Gupta-Allan (MYEGA, Equation (3)) [22] formulas:

$$\tau_\alpha(T) = \tau_\infty \exp\left(\frac{B}{T-T_V}\right), \tag{2}$$

$$\tau_\alpha(T) = \tau_\infty \exp\left(\frac{K}{T}\exp\left(\frac{C}{T}\right)\right), \tag{3}$$

where: $\tau_\infty$ is relaxation time at $T \to \infty$, $T_V$ is Vogel temperature where $\tau_\alpha \to \infty$, and $B$, $K$, $C$ are proportional to activation barriers. The Arrhenius formula corresponds to $T_V = 0$ or $C = 0$; in that case $B$, $K$ are equal to an activation energy divided by the gas constant $R$. The temperature dependence of $\tau_\alpha$ on approaching $T_g$ is commonly characterized by the fragility index $m_f$ (Equation (4)) [21]:

$$m_f = \left.\frac{\mathrm{d}\log_{10}\tau_\alpha(T)}{\mathrm{d}(T_g/T)}\right|_{T=T_g}, \tag{4}$$

where $m_f \approx 16$ indicates a strong glass-former (Arrhenius dependence) and $m_f \approx 200$ indicates a very fragile glass-former (super-Arrhenius dependence). A relationship between fragility and crystallization tendency has been proposed for metallic glass-formers [23], but its applicability to molecular glass-formers is not established.

The secondary β-relaxation may involve local motions of the entire molecule, as in the Johari–Goldstein (JG) process, or intramolecular motions [24]. For a genuine JG process, the relationship between parameter $\tau_\infty$ and activation energy $E_\beta$ of the β-process, shape parameter $a'$ of the α process, and $T_g$ is described by Equation (5) [25]:

$$\frac{E_\beta}{RT_g} = 2.303(2 - 13.7a' - \log_{10}\tau_\infty). \tag{5}$$

The stretching exponent of the Kohlrausch-Williams-Watts function can be estimated from the Cole-Cole shape parameter using the relationship $a' = a^{0.81}$ [26]. This comparison can support the assignment of the β-relaxation but does not uniquely identify its molecular origin.

The investigated compound, 4-cyano-3-fluorophenyl 4-butylbenzoate (Figure 1), denoted as 4CFPB [27], exhibits a nematic phase with a long-range orientational order of long molecular axes [28]. Bulk 4CFPB forms a nematic glass upon cooling, with $T_g \approx 217$ K, and undergoes cold crystallization during subsequent heating [15,16,27]. Two α-relaxations were identified: the lower-frequency $\alpha_1$ process, assigned to rotations around the short molecular axis (s-process), and the higher-frequency $\alpha_2$ process, assigned to molecular precession. The secondary β-relaxation is interpreted as tumbling of molecules [27]. 4CFPB constrained within very small cylindrical nanopores of 6 and 8 nm diameters shows also two α-relaxations, although both have an Arrhenius dependence, instead of super-Arrhenius reported in bulk, and the glass transition temperature is lower, 197 and 207 K for 6 and 8 nm, respectively. The β-relaxation was not resolved in these pores and cold crystallization during heating was hindered [15,16]. These confinement effects were previously discussed in terms of an effective negative pressure [16]. In this work, we investigate the molecular dynamics and glass-forming behavior of 4CFPB confined in anodic aluminum oxide (AAO) cylindrical nanopores with diameters of 35 and 73 nm. The temperature dependence of primary α-relaxations, secondary β-relaxation, $T_g$, and cold crystallization are examined and compared with previous results for bulk 4CFPB and 6 and 8 nm pores.

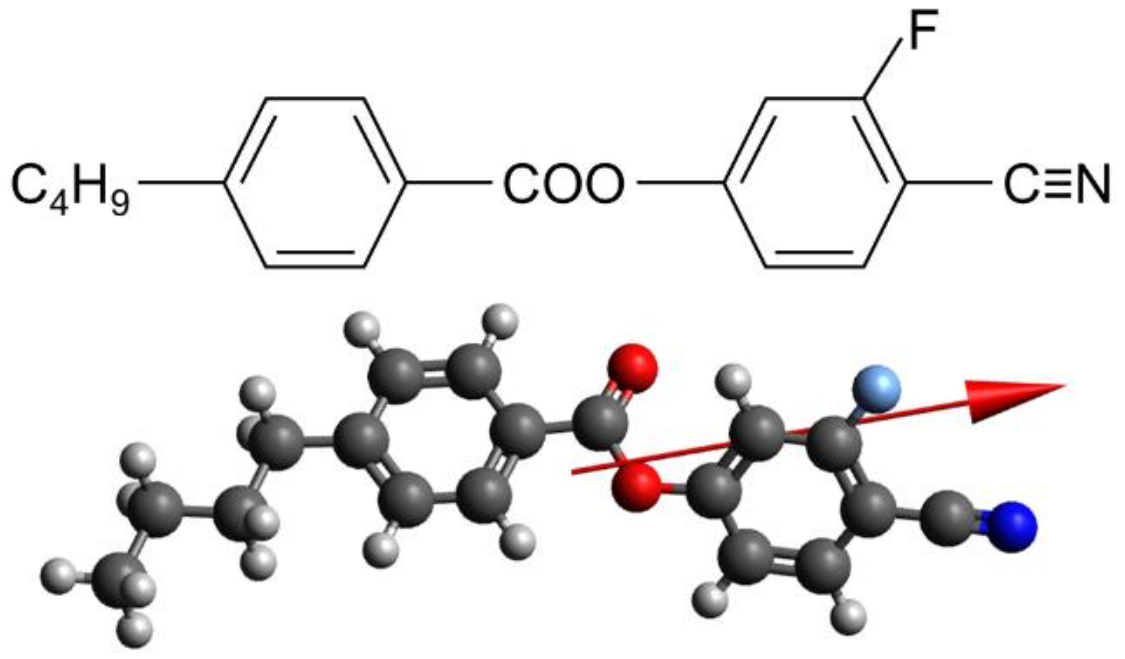

Figure 1. Structural formula of 4CFPB (top) and a molecular model optimized by DFT B3LYP-D3(BJ)/def2TZVPP calculations with a dipole moment of 9.0 D (bottom).

**2. Experimental and computational details**

The 4-cyano-3-fluorophenyl 4-butylbenzoate was synthesized at the Institute of Chemistry, Military University of Technology, Warsaw, Poland. AAO nanoporous matrices with nominal pore diameters of 35 and 73 nm were used to investigate the influence of geometrical confinement on the molecular dynamics of 4CFPB. Prior to infiltration, the empty nanoporous matrices were thermally treated at 373 K for 24 h to remove adsorbed moisture and volatile contaminants, including possible residues from the synthesis of the porous material. Subsequently, 4CFPB was introduced into the nanopores in the isotropic liquid state. Infiltration was carried out for 24 h through spontaneous capillary filling, driven by capillary forces arising from the wetting of the pore walls by the liquid sample. The material was maintained in the isotropic state throughout the infiltration process to facilitate its penetration into the nanoporous structure.

Broadband dielectric spectroscopy (BDS) measurements were performed using a Novocontrol Alpha analyzer over the frequency range of 0.1 Hz-10 MHz during a cooling-heating cycle between 173 and 299 K. The samples were placed between gold electrodes with a sample thickness of 50 μm. The dielectric spectra were analyzed using OriginPro.

Density functional theory (DFT) calculations were performed using the B3LYP-D3(BJ) exchange-correlation functional [29-31] and def2TZVPP basis set [32] as implemented in Gaussian 16, Revision C.01 [33]. Molecular structure was visualized using Avogadro [34].

## 3. Results and discussion

Five dielectric relaxation processes were identified in 4CFPB confined in 35 and 73 nm pores. In order of increasing frequency, these are the Maxwell-Wagner-Sillars (MWS) process, a process tentatively assigned to motions of molecules next to pore walls, and the $\alpha_1$, $\alpha_2$, and β-relaxations. The assignment of these processes is based on previous dielectric studies of bulk and nanoconfined 4CFPB [15,16,27]. Representative fits of the dielectric spectra using Equation (1) are shown in Figures 2 and 3. The resulting dielectric strengths and relaxation times are presented in Figures 4 and 5, respectively. Figure 4 shows additionally the temperature dependence of the dielectric permittivity at 1 kHz, highlighting the isotropic-nematic (Iso/N) transition.

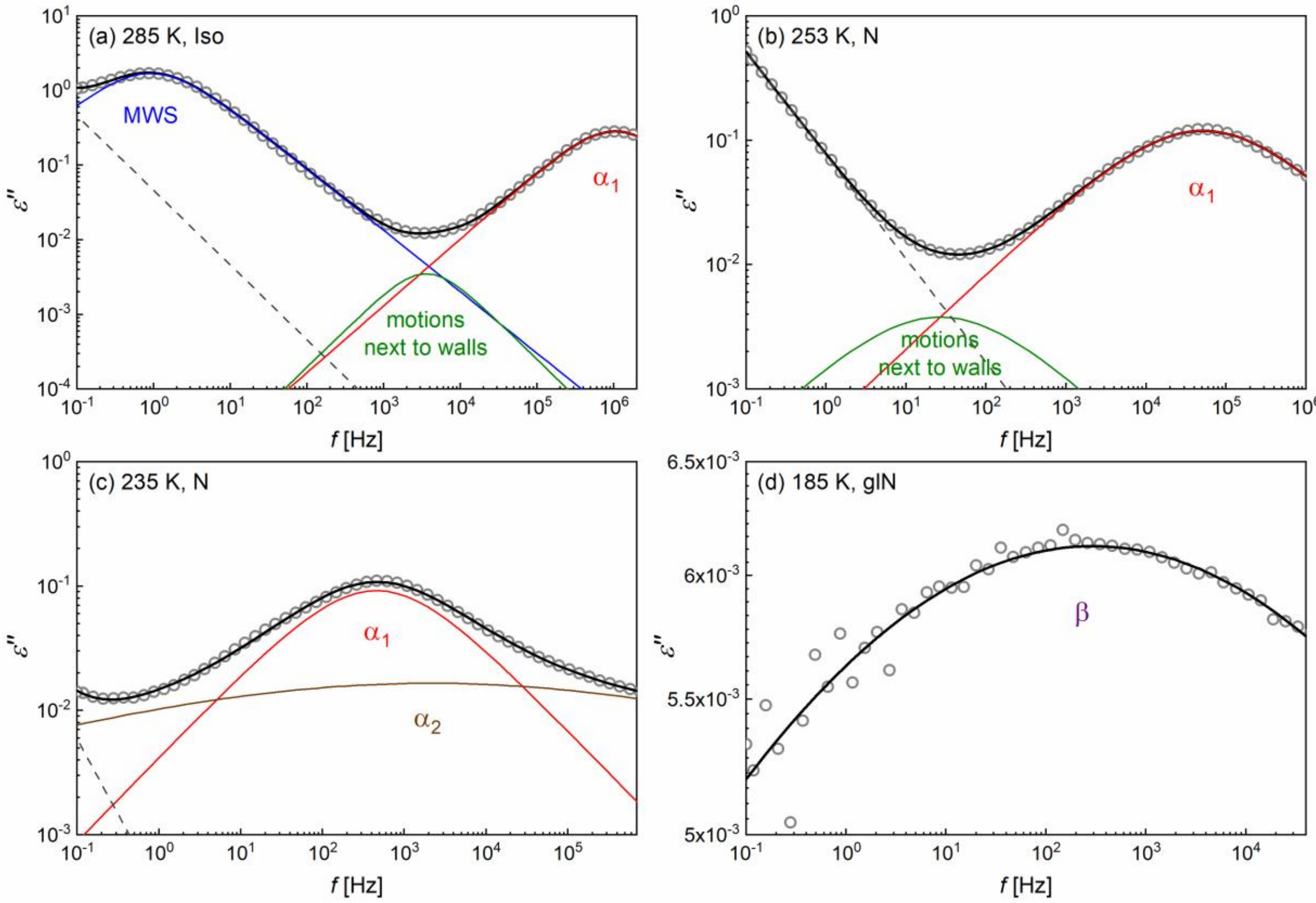

Figure 2. Representative fits of Equation (1) to BDS spectra of 4CFPB in nanopores with 73 nm diameter collected on cooling at 285 K (a), 253 K (b), 235 K (c), and 185 K (d). Points – experimental spectra, solid line – fits of each process, dashed line – background, thick solid line – total fit.

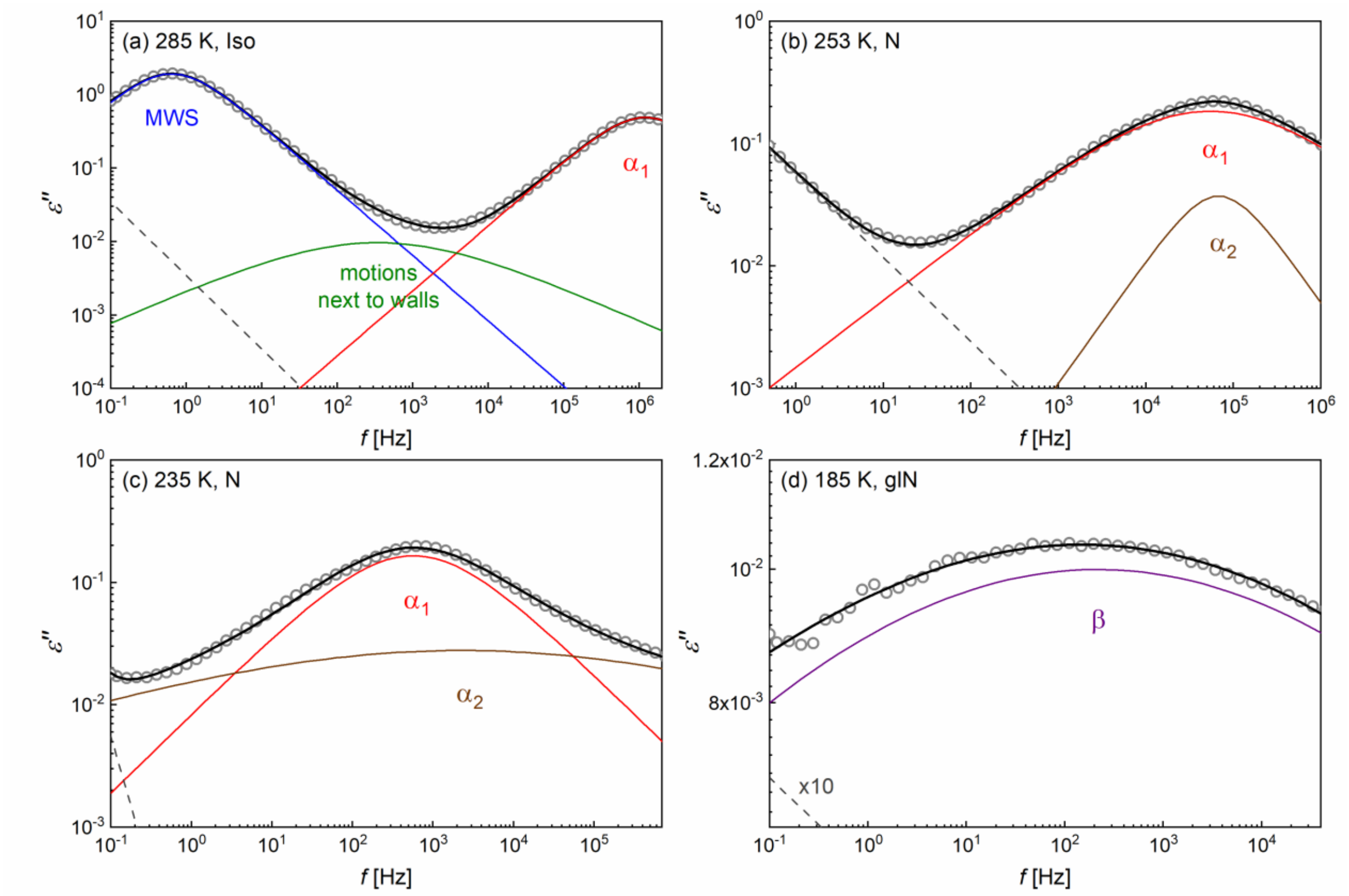

Figure 3. Representative fits of Equation (1) to BDS spectra of 4CFPB in nanopores with 35 nm diameter collected on cooling at 285 K (a), 253 K (b), 235 K (c), and 185 K (d). Points – experimental spectra, solid line – fits of each process, dashed line – background, thick solid line – total fit.

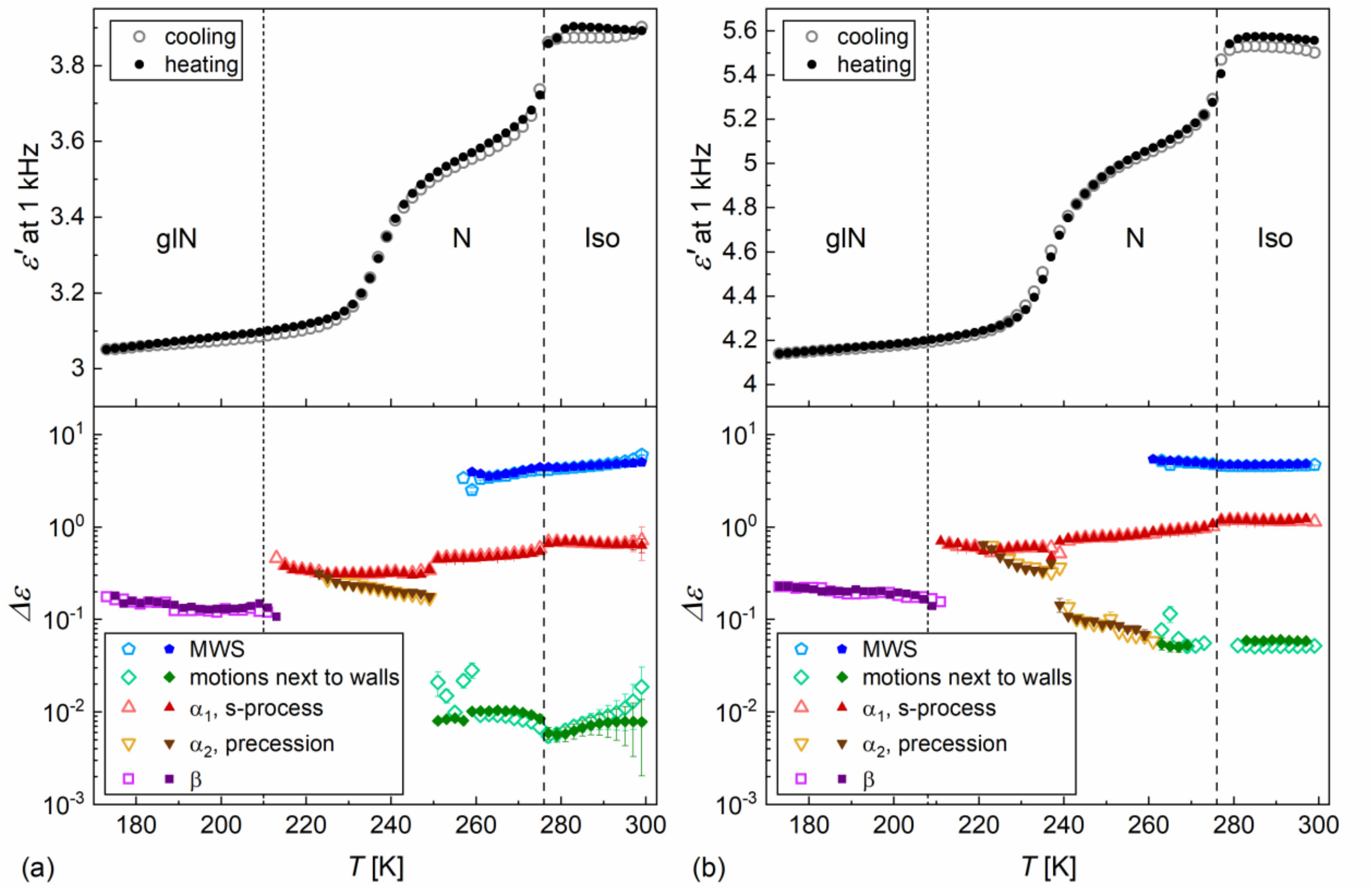

Figure 4. Dielectric dispersion at 1 kHz (upper panels) and dielectric strength of relaxation processes (bottom panels) for 4CFPB in nanopores with 73 nm (a) and 35 nm (b) diameters. Open symbols – cooling, solid symbols – heating.

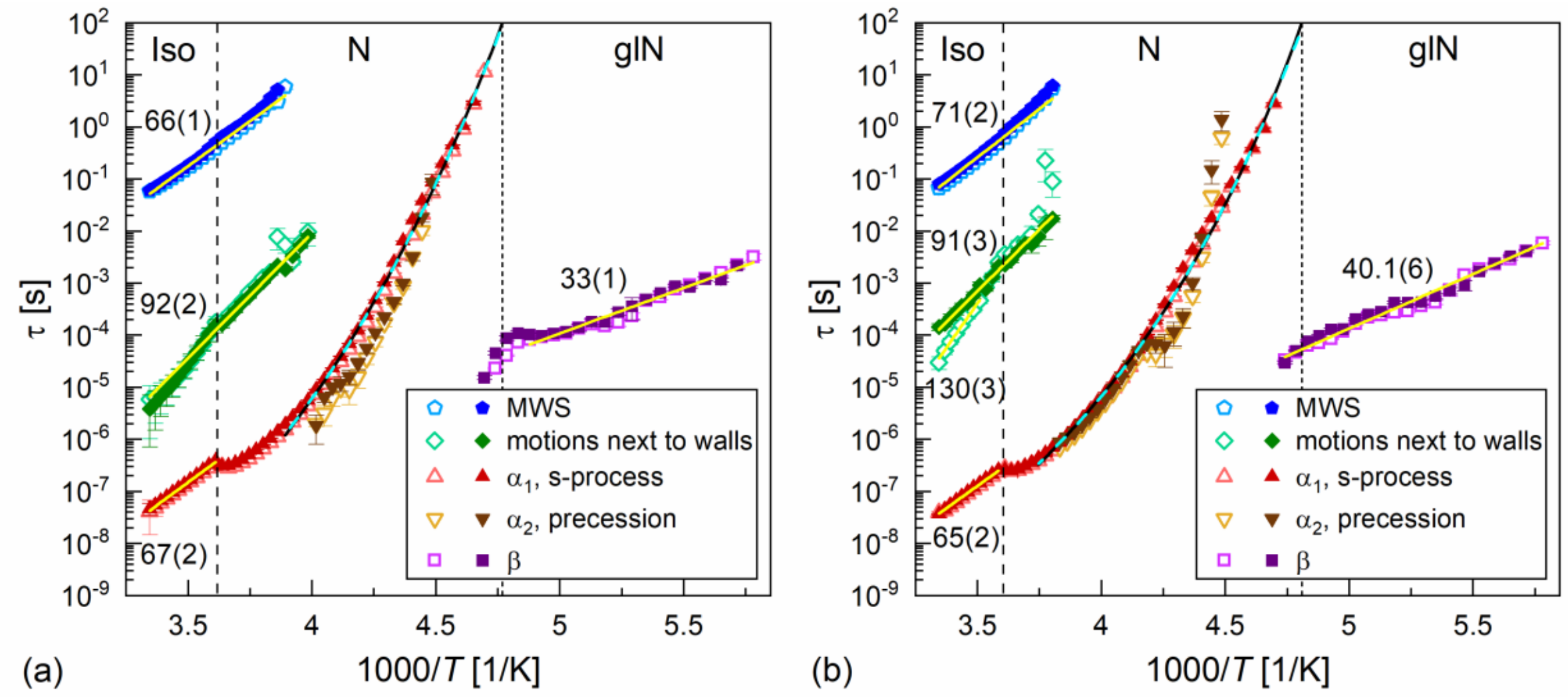

Figure 5. Activation plots of 4CFPB in nanopores with 73 nm (a) and 35 nm (b) diameters. Open symbols – cooling, solid symbols – heating. Light solid line – Arrhenius formula, dark solid line – VFT formula, light dashed line – MYEGA formula.

Maxwell-Wagner-Sillars (MWS) polarization arises from interfacial charge accumulation in heterogeneous systems [35]. The MWS process exhibits Arrhenius behavior with activation energies of 66(1) and 71(2) kJ/mol for 73 and 35 nm pores, respectively (numbers in parentheses are uncertainties at the last digit). The dielectric strength of MWS shows no distinct change at the Iso/N transition in both cases. Upon further cooling in the nematic phase, the MWS relaxation shifts towards lower frequencies, contributing to the decrease in the dielectric response measured at a fixed frequency. The MWS process exhibits relatively weak broadening, with a Cole-Cole shape parameter of approximately 0.1-0.2.

A weak relaxation process is observed between the MWS and $\alpha_l$ processes. It is tentatively assigned to molecular motions close to the core walls [36] because it is present only in nanoconfined 4CFPB samples [15]. Its relaxation time follows Arrhenius behavior, with activation energies of 92(2) and 91(3) kJ/mol for 73 and 35 nm pores, respectively. A higher activation energy of approximately 130 kJ/mol was obtained during heating for the sample confined in 35 nm pores. The strength of this process is almost one order of magnitude smaller in 73 nm than in 35 nm pores. This process was fitted as a Debye model ($a$ = 0) in most of spectra for 73 nm pores, while it was significantly broadened in 35 nm pores, with $a$ ~ 0.55. However, because this contribution is considerably weaker than the neighboring MWS and $\alpha_l$ processes and does not form a clearly resolved loss peak, there is a considerable uncertainty of the shape parameter.

The main $\alpha_l$-relaxation (s-process) has an Arrhenius behavior in isotropic liquid, with activation energies of 67(2) and 65(2) kJ/mol for 73 and 35 nm. There is a decrease of strength of this process at the Iso/N transition, which contributes to an abrupt decrease of $\varepsilon'$. In the nematic phase, the temperature dependence of the $\alpha_l$-relaxation time becomes super-Arrhenius. The fitting results of VFT (Equation (2)) and MYEGA (Equation (3)) models are presented in Table 1. Both models yield slightly lower $T_g$ values for 35 nm pores than for 73 nm pores. The fragility index is also slightly higher for 35 nm pores, particularly according to the VFT model. However, the differences are small, and the overall relaxation behavior is similar for both pore diameters. The shape parameter $a$ is equal to ~0.1 in the isotropic liquid and increases in the nematic phase to ~0.3-0.4 for 73 nm and ~0.3-0.5 for 35 nm pores. Symmetric broadening of the $\alpha_l$-relaxation peak indicates increasing short-time correlations and decreasing long-time correlations of molecular motions [37].

Table 1. Parameters obtained by fitting the VFT (upper part) and MYEGA (lower part) models to the temperature dependence of the $\alpha_l$-relaxation time of 4CFPB confined in 35 and 73 nm pores.

| diameter [nm] | $\log_{10}(\tau_\infty/s)$ | $B$ [K] | $T_V$ [K] | $T_g$ [K] | $m_f$ |
|---|---|---|---|---|---|
| 73 | −16.1(8) | 2540(260) | 149(4) | 209.8(1) | 62(1) |
| 35 | −13.6(3) | 1768(83) | 159(2) | 208.0(1) | 65.7(9) |
| diameter [nm] | $\log_{10}(\tau_\infty/s)$ | $K$ [K] | $C$ [K] | $T_g$ [K] | $m_f$ |
| 73 | −12.4(6) | 278(58) | 674(30) | 210(1) | 60.6(5) |
| 35 | −10.6(2) | 129(12) | 798(14) | 207.6(5) | 61.3(2) |

In 73 nm pores, the $\alpha_2$-relaxation, assigned to molecular precession, becomes resolved on the high-frequency side of the $\alpha_1$ process below 249 K. It exhibits considerable broadening, with a Cole-Cole shape parameter of approximately 0.7-0.9. Meanwhile, it shows as a narrower process with $a \sim 0$ and relaxation time almost overlapping with the $\alpha_1$-relaxation for a sample within 35 nm pores in 239-261 K. Below 239 K, the $\alpha_2$-relaxation broadens to $a \sim$ 0.7-0.9 and behaves similarly as in 73 nm pores. The temperature dependence of the $\alpha_2$-relaxation time was not fitted with VFT and MYEGA formulas due to a smaller number and larger scatter of experimental points than for $\alpha_1$.

The β-relaxation time is a very wide process with $a \sim$ 0.8-0.9 in both samples and follows the Arrhenius behavior. The activation energy is equal to 33(1) and 40.1(6) kJ/mol for 73 and 35 nm. To check if the observed β-relaxation is the genuine JG process, Equation (5) was filled with experimental parameters. The results of the MYEGA model were assumed for $\alpha_1$-relaxation The left-hand side is equal to 19.1(6) and 23.2(4) for 73 and 35 nm, while the right-hand side is equal to 16.6(8) and 22.3(5) for 73 and 35 nm. Small differences between left and right-hand sides of Equation (5) indicate that the β-relaxation may be the genuine JG process and originate from tumbling of whole molecules. An alternative interpretation are intra-molecular rotations within the COO bridge between aromatic rings. The rotation around the C-(C=O) bond has a calculated energy barrier of 30.6 kJ/mol (Figure 6a), which is close to the experimental activation energy of the β-process for 73 nm. It can be coupled with rotations around the O-C bond, with a lower calculated energy barrier of 1.7-2.7 kJ/mol (Figure 6b). Higher activation energy in 35 nm pores is probably an effect of a stronger confinement, especially that the β process is completely suppressed in very small nanopores of 6 and 8 nm [15,16].

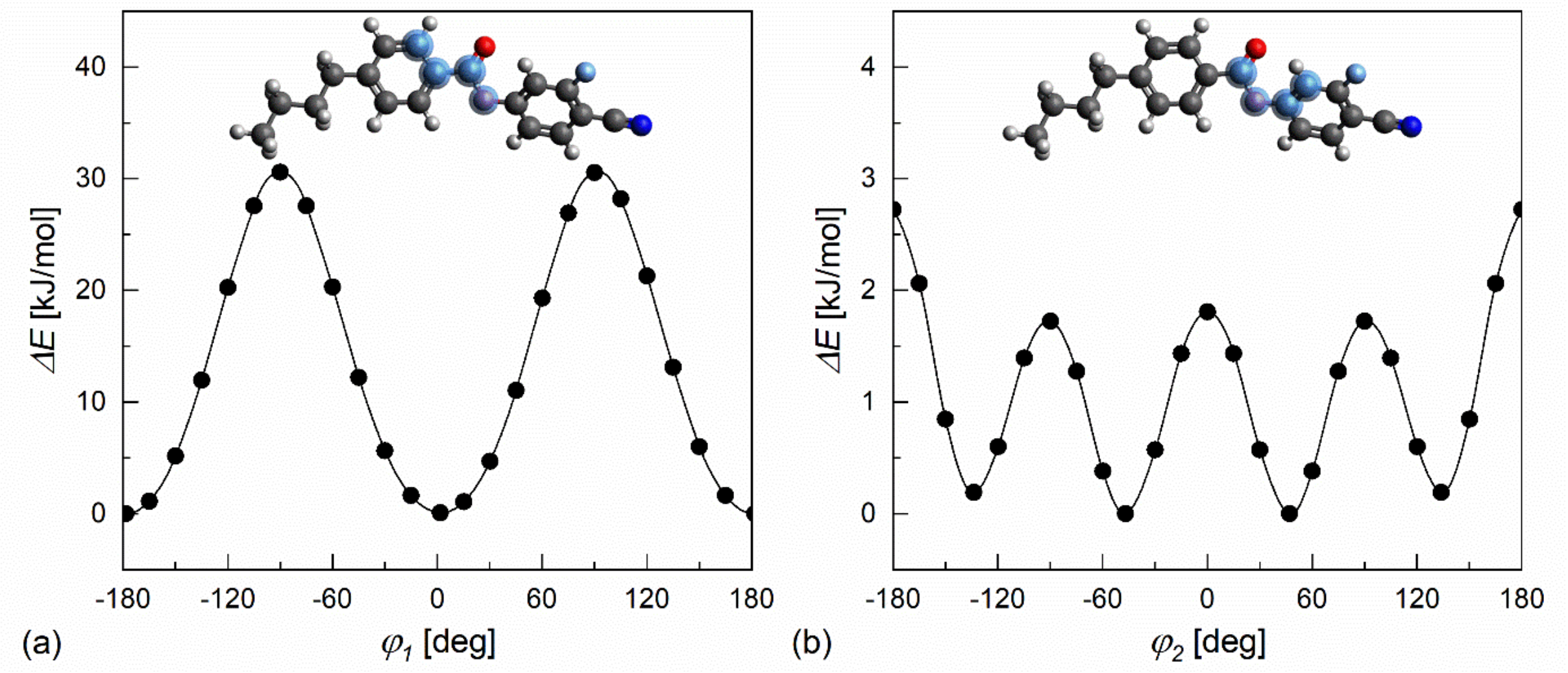

Figure 6. Energy scans calculated by DFT at the B3LYP-D3(BJ)/def2TZVPP level for torsional angles C-C-C(=O)-O, $\varphi_1$ (a) and C(=O)-O-C-C, $\varphi_2$ (b) in the 4CFPB molecule.

The glass transition temperature decreases with decreasing pore diameter, from approximately 217 K in bulk 4CFPB to 210 K in 73 nm pores, 208 K in 35 nm pores, 207 K in 8 nm pores, and 197 K in 6 nm pores [15,16,27]. The $m_f$ fragility index of the $\alpha_1$-relaxation time is ~60 both in bulk [27] and in 73, 35 nm pores. Only very small pores of 8, 6 nm lead to a significant decrease in $m_f$ to ca. 16, which corresponds to the Arrhenius behavior of the $\alpha_1$-relaxation time [15,16]. The cold crystallization is suppressed in 73, 35 nm pores. The decrease of $\alpha_1$ process on heating in 8, 6 nm pores is interpreted as aligning of molecules along the pore walls, instead of cold crystallization [15]. The aligning is indicated also by small-angle neutron scattering for 18 nm pores, while results for 35 nm pores are ambiguous [38]. Same strength of $\alpha_1$ on cooling and heating in 73, 35 nm pores suggests that aligning at walls is weaker in wider pores. Restricted molecular transport under confinement may hinder crystal growth and contribute to the absence of detectable cold crystallization during heating. It is in agreement with study of cold crystallization of bulk 4CFPB, which was dependent mainly on the diffusion rate at slow cooling [27], which is also the case for BDS measurements presented here.

## 4. Summary and conclusions

The molecular dynamics and glass-forming behavior of 4CFPB confined in AAO cylindrical nanopores with diameters of 35 and 73 nm were investigated using broadband dielectric spectroscopy and compared with previous results for bulk 4CFPB and confinement in 6 and 8 nm pores. The glass transition temperature decreases from approximately 217 K in bulk 4CFPB to 210 K and 208 K in 73 and 35 nm pores, respectively. However, the $\alpha_1$-relaxation, corresponding to the molecular motions around the short molecular axes, retains its super-Arrhenius temperature dependence, with a fragility index close to 60, similar to that of the bulk material. This contrasts with the Arrhenius behavior previously observed in 6 and 8 nm pores. Thus, the reduction in $T_g$ under confinement is not necessarily accompanied by a substantial change in fragility. The secondary β-relaxation remains detectable in both 35 and 73 nm pores, with activation energies of 40.1(6) and 33(1) kJ/mol, respectively, whereas it was not resolved in previously studied 6 and 8 nm pores. Comparison with the Johari-Goldstein criterion and DFT-calculated torsional barriers suggests possible contributions from both whole-molecule and intramolecular motions, but does not provide an unambiguous assignment of the relaxation mechanism. Cold crystallization was not observed during heating in either pore size, despite the relatively small changes in fragility compared with bulk 4CFPB. Restricted molecular transport contributes probably to this behavior, although the mechanism was not directly investigated.

Overall, the results show that confinement affects the glass transition temperature, relaxation dynamics, and crystallization of 4CFPB to different extents. In particular, the comparison across pore sizes demonstrates that a decrease in $T_g$ can occur without a transition from super-Arrhenius to Arrhenius relaxation behavior.

**Acknowledgements:** We thank Prof. Maria Massalska-Arodź from the Henryk Niewodniczański Institute of Nuclear Physics Polish Academy of Sciences for valuable remarks regarding the manuscript. We gratefully acknowledge Polish high-performance computing infrastructure PLGrid (HPC Center: ACK Cyfronet AGH) for providing computer facilities and support within computational grant no. PLG/2026/019109.

**Declaration statement:** The authors declare no conflict of interest.

**References:**

[1] M. Żak, J. Herman, P. Kula, Deuterium reinforced liquid crystals – synthesis and photochemical stability of deuterium labelled cyano-oligophenyls, J. Mol. Liq. 435 (2025) 128096, https://doi.org/10.1016/j.molliq.2025.128096.

[2] E. Wojda, M. Zając, P. Perkowski, M. Urbańska, Synthesis, Properties, and Enantioseparation of Four-Ring Racemic Smectics, Materials 19 (2026) 1719; https://doi.org/10.3390/ma19091719.

[3] N. Podoliak, V. Novotná, T. Jurkovičová, V. Hamplová, D. Pociecha, M. Cigl, Liquid crystalline derivatives exhibiting smectic phases with ferro- and antiferro-electric properties, Soft Matter 22 (2026) 2389-2398, https://doi.org/10.1039/d6sm00028b.

[4] N. Kapernaum, F. Knecht, C.S. Hartley, J.C. Roberts, R.P. Lemieux, F. Giesselmann, Formation of smectic phases in binary liquid crystal mixtures with a huge length ratio, Beilstein J. Org. Chem. 8 (2012) 1118-1125, https://doi.org/10.3762/bjoc.8.124.

[5] B. Loska, Y. Arakawa, K. Merkel, Mix and match: twist-bend nematic behaviour in liquid crystal dimer mixtures of the CBOnOCB, Liq. Cryst. 52 (2025) 1507-1523, https://doi.org/10.1080/02678292.2025.2517869.

[6] M.C. Mora, J. Brenes, C. Camacho, E. Castellón, The Role of Functional Groups in Substituted Benzoic Acids Used as Dopants in Liquid Crystal Mixtures on the Nematic-Isotropic Transitions, J. Phys. Chem. B 130 (2026) 7250-7261, https://doi.org/10.1021/acs.jpcb.6c01531.

[7] S. Lalik, O. Stefańczyk, N. Górska, K. Kumar, S.-i. Ohkoshi, M. Marzec, Modifications of EHPDB Physical Properties through Doping with $Fe_2O_3$ Nanoparticles (Part II), Int. J. Mol. Sci. 23 (2022) 50, https://doi.org/10.3390/ijms23010050.

[8] A.S. Mera, Y. Boussoualem, B. Duponchel, Y. Lin, J.-F. Blach, S. Saitzek, C. Davoisne, R. Douali, A. Daoudi, Ferroelectric and dielectric studies of a ferroelectric liquid crystal (FLC) dispersed ferroelectric nanoparticles (NPs): Impact of NP dispersions on the rotational viscoelastic properties of the FLC, J. Mol. Liq. 448 (2026) 129363, https://doi.org/10.1016/j.molliq.2026.129363.

[9] B. Pratap Singh, S. Sikarwar, K. Kumar Pandey, R. Manohar, M. Depriester, D. Pratap Singh, Carbon Nanotubes Blended Nematic Liquid Crystal for Display and Electro-Optical Applications, Electron. Mater. 2 (2021) 466-481, https://doi.org/10.3390/electronicmat2040032.

[10] L. Lisetski, L. Bulavin, N. Lebovka, Effects of Dispersed Carbon Nanotubes and Emerging Supramolecular Structures on Phase Transitions in Liquid Crystals: Physico-Chemical Aspects, Liquids 3 (2023) 246-277, https://doi.org/10.3390/liquids3020017.

[11] A. Drzewicz, Sz. Folta, O. Adamczyk, M. Żak, E. Juszyńska-Gałązka, Isotopic and Confinement Effects on the Phase Behavior and Molecular Dynamics of Protonated and Deuterated Terphenyls in Electrospun Fibers, J. Phys. Chem. B 130 (2026) 2676-2688, https://doi.org/10.1021/acs.jpcb.5c07839.

[12] M. Busch, A.V. Kityk, W. Piecek, T. Hofmann, D. Wallacher, S. Całus, P. Kula, M. Steinhart, M. Eich, P. Huber, A ferroelectric liquid crystal confined in cylindrical nanopores: reversible smectic layer buckling, enhanced light rotation and extremely fast electro-optically active Goldstone excitations, Nanoscale 9 (2017) 19086-19099, https://doi.org/10.1039/c7nr07273b.

[13] Z. Li, K. Sentker, M. Brinker, M. Dimitrova Lippmann, O.H. Seeck, A.V. Kityk, B.M. Ocko, P. Huber, Directed Self-Assembly of the Organic Semiconductor C8-BTBT-C8 in Anodic Aluminum Oxide Nanopores, ACS Appl. Nano Mater. 9 (2026) 273-282, https://doi.org/10.1021/acsanm.5c04558.

[14] A.Z. Maksym, A.S. Andrushchak, Y. Shchur, B. Sahraoui, P. Kula, M. Lelonek, M. Busch, P. Huber, A.V. Kityk, Self-Assembly of Bent-Core Nematics in Nanopores, Small 21 (2025) e06651, https://doi.org/10.1002/smll.202506651.
[15] T. Rozwadowski, M. Massalska-Arodź, J. Krawczyk, E. Juszyńska-Gałązka, Molecular dynamics of 4-cyano-3-fluorophenyl 4-butylbenzoate (4CFPB) glass-forming liquid crystal in unidirectional silicon nanopores, Liq. Cryst. 41 (2014) 1073-1079, https://doi.org/10.1080/02678292.2014.900706.
[16] T. Rozwadowski, M. Massalska-Arodź, M. Jasiurkowska-Delaporte, Negative pressure effects on molecular dynamics and phase diagram of glass-forming nematic liquid crystal 4-cyano-3-fluorophenyl 4-butylbenzoate (4CFPB) confined in nanopores, J. Mol. Liq. 279 (2019) 127-132, https://doi.org/10.1016/j.molliq.2019.01.106.
[17] A. Gujral, L. Yu, M.D. Ediger, Anisotropic organic glasses, Curr. Opin. Solid State Mater. Sci. 22 (2018) 49-57, https://doi.org/10.1016/j.cossms.2017.11.001.
[18] V. Rudenko, A. Tolochko, S. Bugaychuk, D. Zhulai, G. Klimusheva, G. Yaremchuk, T. Mirnaya, Y. Garbovskiy, Glass-Forming Ionic Liquid Crystal Gold-Carbon Nanocomposites with Ultrafast Optical Nonlinearity Sign Reversal, J. Compos. Sci. 9 (2025) 472, https://doi.org/10.3390/jcs9090472.
[19] F. Kremer, A. Schönhals (Eds.), Broadband Dielectric Spectroscopy, Springer-Verlag, Berlin Heidelberg 2003, https://doi.org/10.1007/978-3-642-56120-7.
[20] K.S. Cole, R.H. Cole, Dispersion and Absorption in Dielectrics I. Alternating Current Characteristics, J. Chem. Phys. 9 (1941) 341-351, https://doi.org/10.1063/1.1750906.
[21] R. Böhmer, K.L. Ngai, C.A. Angell, D.J. Plazek, Nonexponential relaxations in strong and fragile glass formers, J. Chem. Phys. 99 (1993) 4201-4209, https://doi.org/10.1063/1.466117.
[22] J.C. Mauro, Y. Yue, A.J. Ellison, P.K. Gupta, D.C. Allan, Viscosity of glass-forming liquids, Proc. Natl. Acad. Sci. 106 (2009) 19780-19784, https://doi.org/10.1073/pnas.0911705106.
[23] H. Tanaka, Relationship among glass-forming ability, fragility, and short-range bond ordering of liquids, J. Non-Cryst. Solids 351 (2005) 678-690, https://doi.org/10.1016/j.jnoncrysol.2005.01.070.
[24] G.P. Johari, M. Goldstein, Viscous Liquids and the Glass Transition. II. Secondary Relaxations in Glasses of Rigid Molecules, J. Chem. Phys. 53 (1970) 2372-2388, https://doi.org/10.1063/1.1674335.
[25] K.L. Ngai, S. Capaccioli, Relation between the activation energy of the Johari-Goldstein β relaxation and $T_g$ of glass formers, Phys. Rev. E 69 (2004) 031501, https://doi.org/10.1103/PhysRevE.69.031501.
[26] F. Alvarez, A. Alegra, J. Colmenero, Relationship between the time-domain Kohlrausch-Williams-Watts and frequency-domain Havriliak-Negami relaxation functions, Phys. Rev. B 44 (1991) 7306-7312, https://doi.org/10.1103/PhysRevB.44.7306.
[27] T. Rozwadowski, M. Massalska-Arodź, Ł. Kolek, K. Grzybowska, A. Bąk, K. Chłędowska, Kinetics of Cold Crystallization of 4-Cyano-3-fluorophenyl 4-Butylbenzoate (4CFPB) Glass Forming Liquid Crystal. I. Nonisothermal Process As Studied by Microscopic, Calorimetric, and Dielectric Methods, Cryst. Growth Des. 15 (2015) 2891-2900, https://doi.org/10.1021/acs.cgd.5b00328.
[28] C. Tschierske (Ed.), Liquid Crystals, Materials Design and Self-Assembly, Springer-Verlag, Berlin Heidelberg 2012.
[29] C. Lee, W. Yang, R.G. Parr, Development of the Colle-Salvetti correlation-energy formula into a functional of the electron density, Phys. Rev. B, 37 (1988) 785-789, https://doi.org/10.1103/PhysRevB.37.785.
[30] A.D. Becke, Density-functional thermochemistry. III. The role of exact exchange, J. Chem. Phys. 98 (1993) 5648-5652, https://doi.org/10.1063/1.464913.
[31] S. Grimme, S. Ehrlich, L. Goerigk, Effect of the damping function in dispersion corrected density functional theory, J. Comput. Chem. 32 (2011) 1456-1465, https://doi.org/10.1002/jcc.21759.
[32] F. Weigend, R. Ahlrichs, Balanced basis sets of split valence, triple zeta valence and quadruple zeta valence quality for H to Rn: Design and assessment of accuracy, Phys. Chem. Chem. Phys. 7 (2005) 3297-3305, https://doi.org/10.1039/B508541A.
[33] Gaussian 16, Revision C.01, M.J. Frisch, G.W. Trucks, H.B. Schlegel, G.E. Scuseria, M.A. Robb, J.R. Cheeseman, G. Scalmani, V. Barone, G.A. Petersson, H. Nakatsuji, X. Li, M. Caricato, A.V. Marenich, J. Bloino, B.G. Janesko, R. Gomperts, B. Mennucci, H.P. Hratchian, J.V. Ortiz, A.F. Izmaylov, J.L. Sonnenberg, D. Williams-Young, F. Ding,

F. Lipparini, F. Egidi, J. Goings, B. Peng, A. Petrone, T. Henderson, D. Ranasinghe, V.G. Zakrzewski, J. Gao, N. Rega, G. Zheng, W. Liang, M. Hada, M. Ehara, K. Toyota, R. Fukuda, J. Hasegawa, M. Ishida, T. Nakajima, Y. Honda, O. Kitao, H. Nakai, T. Vreven, K. Throssell, J.A. Montgomery, Jr., J.E. Peralta, F. Ogliaro, M.J. Bearpark, J.J. Heyd, E.N. Brothers, K.N. Kudin, V.N. Staroverov, T.A. Keith, R. Kobayashi, J. Normand, K. Raghavachari, A.P. Rendell, J.C. Burant, S.S. Iyengar, J. Tomasi, M. Cossi, J.M. Millam, M. Klene, C. Adamo, R. Cammi, J.W. Ochterski, R.L. Martin, K. Morokuma, O. Farkas, J.B. Foresman, D.J. Fox, Gaussian, Inc., Wallingford CT, 2019.

[34] M.D. Hanwell, D.E. Curtis, D.C. Lonie, T. Vandermeersch, E. Zurek, G. R. Hutchison, Avogadro: an advanced semantic chemical editor, visualization, and analysis platform, J. Cheminf. 4 (2012) 17, https://doi.org/10.1186/1758-2946-4-17.

[35] M. Samet, V. Levchenko, G. Boiteux, G. Seytre, A. Kallel, A. Serghei, Electrode polarization vs. Maxwell-Wagner-Sillars interfacial polarization in dielectric spectra of materials: Characteristic frequencies and scaling laws, J. Chem. Phys. 142 (2015) 194703, https://doi.org/10.1063/1.4919877.

[36] M. Arndt, R. Stannarius, W. Gorbatschow, F. Kremer, Dielectric investigations of the dynamic glass transition in nanopores, Phys. Rev. E 54 (1996) 5377-5390, https://doi.org/10.1103/PhysRevE.54.5377.

[37] A.K. Jonscher, A new understanding of the dielectric relaxation of solids, J. Mater. Sci. 16 (1981) 2037-2060, https://doi.org/10.1007/BF00542364.

[38] E. Juszyńska-Gałązka, W. Zając, Small angle neutron scattering on n-cyano-3 fluorophenyl 4-butylbenzoates (n = 4,6,8) liquid crystals in mesoporous AAO membranes. A feasibility study, Phase Trans. 96 (2023) 89-96, https://doi.org/10.1080/01411594.2023.2167656.